\documentclass[preprint,aps,prd,superscriptaddress,preprintnumbers,amsmath,amssymb]{revtex4}

\newcommand{\fft}[2]{{\frac{#1}{#2}}}
\newcommand{\ft}[2]{{\textstyle\frac{#1}{#2}}}

\begin{document}
\preprint{MCTP-10-57}

\title{Holographic $c$-theorems and higher derivative gravity}

\author{James T.~Liu}
\email{jimliu@umich.edu}
\affiliation{Michigan Center for Theoretical Physics,
Randall Laboratory of Physics,
The University of Michigan,
Ann Arbor, MI 48109--1040, USA}

\author{Wafic Sabra}
\email{ws00@aub.edu.lb}
\affiliation{Centre for Advanced Mathematical Sciences and Physics Department,
American University of Beirut,
Lebanon}

\author{Zhichen Zhao}
\email{zhichen@umich.edu}
\affiliation{Michigan Center for Theoretical Physics,
Randall Laboratory of Physics,
The University of Michigan,
Ann Arbor, MI 48109--1040, USA}

\begin{abstract}

In AdS/CFT, the holographic Weyl anomaly computation relates the $a$-anomaly
coefficient to the properties of the bulk action at the UV fixed
point.  This universal behavior suggests the possibility of a holographic
$c$-theorem for the $a$-anomaly under flows to the IR.  We prove such a
$c$-theorem for higher curvature Lovelock gravity, where the bulk equations
of motion remain second order.  We also explore $f(R)$ gravity as a toy
model where higher derivatives cannot be avoided.  In this case, monoticity
of the flow requires an additional condition related to the higher derivative
nature of the theory.  This is in contrast to the case of $f(R)$ black hole
entropy, where the second law follows from application of the full
Einstein equations and the null energy condition.

\end{abstract}

\maketitle

\section{Introduction}

Central charges in conformal field theories can often be thought of as
a proxy for the number of degrees of freedom exhibited by the theory.  In
this context, the Zamolodchikov $c$-theorem \cite{Zamolodchikov:1986gt}
is a powerful result for
two-dimensional conformal field theories.  It states that there exists a
$c$-function which is monotonically decreasing along flows from the UV to
IR, and which is equal to the central charge at the fixed points of the flow.
This is a direct indication that UV degrees of freedom of the CFT are
removed as the theory flows to the IR.

While two dimensional conformal field theories are rather special, there
have been numerous attempts to generalize the $c$-theorem to higher
dimensions.  However, one obstacle that needs to be surmounted in doing so
is the realization that there may be multiple candidates for a satisfactory
$c$-function.  For example, in four dimensions, the Weyl anomaly has the
well known form
\begin{equation}
\langle T^\mu_\mu\rangle=\fft{c}{16\pi^2}C_{\mu\nu\rho\sigma}^2
-\fft{a}{16\pi^2}E_4,
\end{equation}
where $E_4=R_{\mu\nu\rho\sigma}^2-4R_{\mu\nu}^2+R^2$ is the four-dimensional
Euler density.  In this context, Cardy demonstrated that, while the
$c$ anomaly coefficient may not be the proper object to investigate,
the $a$ coefficient
appears to have the desired monotonicity property along flows
\cite{Cardy:1988cwa}.  This has subsequently been confirmed in various
situations \cite{Osborn:1989td,Jack:1990eb,Anselmi:1997am,Anselmi:1997ys},
and further investigated in the context of $a$-maximization
\cite{Intriligator:2003jj,Kutasov:2003iy,Kutasov:2003ux,Barnes:2004jj}.

While the above investigations have been carried out in a field theory
context, AdS/CFT allows the possibility of a holographic version of the
$c$-theorem.  At large $N$, the $a$ and $c$ anomalies are equal, and can
be computed holographically; for $\mathcal N=4$ super-Yang-Mills, the
result is simply $a=c=N^2/4$ \cite{Henningson:1998gx}.  This result may
be extended to renormalization group flows, which correspond to radial flow
in the bulk dual.  A $c$-theorem can then be proven at leading order in
large $N$ by examining the flow equations for domain wall solutions
interpolating between the UV and IR
\cite{Alvarez:1998wr,Girardello:1998pd,Freedman:1999gp,Sahakian:1999bd}.

Recently, the leading order holographic $c$-theorem has been extended to the
case where the bulk theory may contain higher order
curvature terms, corresponding to moving away from the leading behavior
in AdS/CFT \cite{Sinha:2010ai,Oliva:2010eb,Myers:2010ru,Oliva:2010zd,Myers:2010xs,Sinha:2010pm,Myers:2010tj}.  In particular, the theories of interest include the addition
of Gauss-Bonnet and `quasi-topological' curvature-cubed terms to the bulk action.
By generalizing
the holographic $a$-anomaly, computed for a general bulk action in
\cite{Imbimbo:1999bj}, Myers and Sinha constructed an appropriate $a$-function that
is monotonic along radial flow \cite{Myers:2010xs,Sinha:2010pm,Myers:2010tj}.
A key element of this holographic $c$-theorem
rests on the fact that the equations of motion contain no higher than second
derivatives of the metric when expanded on the AdS background.  Gauss-Bonnet
gravity manifestly satisfies this requirement, as does the quasi-topological
theory, which was explicitly constructed to have this property \cite{Oliva:2010eb,Myers:2010ru}.

Here we extend the investigation of Myers and Sinha in two directions.
Firstly, we prove a holographic $c$-theorem for the case of Lovelock bulk
theories.  These theories generalize Gauss-Bonnet gravity in $d+1$
dimensions by the addition of $d'$ dimensional Euler densities where
$d'<d$.  While this result is mostly formal, in the sense that higher
Lovelock terms are only present in theories in dimensions too large for
practical applications, it nevertheless suggests that the holographic
$c$-theorem extends to arbitrary orders in the bulk curvature, so long
as higher derivative terms in the equations of motion are able to be
controlled.

Secondly, we address what happens when the bulk action is not restricted
to second order equations of motion by examining $f(R)$ gravity as a toy
model.  In this case, we find that the constructed $a$-function may
deviate from monotonicity by a term that is explicitly of higher derivative
order.  We suggest that this term, which may lead to a violation of the
holographic $c$-theorem, is related to the additional ghost modes of the
theory.  From an AdS/CFT point of view, this would correspond to a breakdown
of unitarity in the dual field theory.

While this work was being prepared, we became aware of 
\cite{Myers:2010tj}, which overlaps with some of our results.

\section{A holographic $c$-theorem for Lovelock gravity}

While higher-curvature gravitational actions generically lead to higher
derivative equations of motion and ensuing pathologies such as ghosts, a
special class of higher-curvature actions may be constructed that
nevertheless give rise to second order equations of motion for the metric.
These are the Lovelock actions, which are constructed out of the
$(d+1)$-dimensional continuation of lower dimensional Euler densities
\cite{Lovelock:1971yv}
\begin{equation}
S=\fft1{2\kappa^2}\int d^{d+1}x\sqrt{-g}\sum_m\alpha_mL^{(m)}
+S_{\mathrm{matter}}.
\label{eq:lovelock}
\end{equation}
Here the $m$-th Lovelock term is the Euler invariant in $2m$ dimensions
\begin{equation}
L^{(m)}=\fft1{2^m}\delta^{a_1b_1\cdots a_mb_m}_{c_1d_1\cdots c_md_m}
R_{a_1b_1}{}^{c_1d_1}\cdots R_{a_mb_m}{}^{c_md_m}.
\end{equation}
In particular, $L^{(0)}=1$ is a cosmological constant, $L^{(1)}=R$ is
the ordinary Einstein-Hilbert term and $L^{(2)}=R_{\mu\nu\rho\sigma}^2
-4R_{\mu\nu}^2+R^2$ is the Gauss-Bonnet invariant.
The equation of motion following from (\ref{eq:lovelock}) is simply
$G_{ab}=\kappa^2T_{ab}$, where the generalized Einstein tensor
is given by $G_{ab}=\sum_m\alpha_mG_{ab}^{(m)}$, with
\begin{equation}
G^{e\,(m)}_f=-\fft1{2^{m+1}}\delta^{ea_1b_1\cdots a_mb_m}_{fc_1d_1\cdots c_md_m}
R_{a_1b_1}{}^{c_1d_1}\cdots R_{a_mb_m}{}^{c_md_m}.
\label{eq:get}
\end{equation}

With a suitable choice of the cosmological constant, we take the bulk Lovelock
action (\ref{eq:lovelock}) to be dual to a $d$-dimensional CFT.  As
demonstrated in \cite{Imbimbo:1999bj}, the $d$-dimensional type A trace
anomaly ({\it i.e.}\ the term proportional to the Euler characteristic) is
universal in holographic renormalization, and its coefficient may be expressed
as
\begin{equation}
a_{\rm UV}=-\fft{\pi^{d/2}}{2\kappa^2}\fft{\ell^{d+1}}{(d/2)!^2}
f(\mathrm{AdS}),
\label{eq:acc}
\end{equation}
where
\begin{equation}
f(\mathrm{AdS})=\sum_m\alpha_mL^{(m)}\Big|_{\mathrm{AdS}}
\label{eq:lovfads}
\end{equation}
is the on-shell Lagrangian evaluated on the asymptotic AdS background with
radius $\ell$.

In order to construct a suitable $a$-function, we need to promote the
$a_{\rm UV}$ central charge (\ref{eq:acc}) at the UV fixed point into a
function $a(r)$ of the radial flow.  Here, for simplicity, we consider a
radial slicing of the bulk space into flat slices of the form
\begin{equation}
ds^2=e^{2A(r)}(-dt^2+d\vec x_{d-1}^2)+dr^2,
\label{eq:slice}
\end{equation}
with curvature components
\begin{equation}
R_{\mu\nu\rho\sigma}=-A'^2(g_{\mu\rho}g_{\nu\sigma}-g_{\mu\sigma}
g_{\nu\rho}),\qquad
R_{\mu r\nu r}=-(A''+A'^2)g_{\mu\nu}.
\label{eq:riemanns}
\end{equation}
This allows us to evaluate the individual Lovelock terms
\begin{equation}
L^{(m)}=(-A'^2)^m\fft{(d+1)!}{(d+1-2m)!}-2mA''(-A'^2)^{m-1}\fft{d!}{(d+1-2m)!}.
\end{equation}
The second term is unimportant at AdS fixed points since we have
\begin{equation}
A\sim r/\ell,\qquad A'\sim1/\ell,\qquad A''\sim0,\qquad\mbox{as}\qquad
r\to\infty.
\end{equation}
As a result, the on-shell value of $L^{(m)}$ takes the form
\begin{equation}
L^{(m)}\Big|_{\mathrm{AdS}}=(-1)^m\fft{(d+1)!}{(d+1-2m)!\ell^{2m}},
\end{equation}
so that the $a_{\rm UV}$ central charge (\ref{eq:acc}) may be expressed as
\begin{equation}
a_{\rm UV}=-\fft{\pi^{d/2}}{2\kappa^2(d/2)!^2}\sum_m\alpha_m
(-1)^m\fft{(d+1)!\ell^{d+1-2m}}{(d+1-2m)!}.
\end{equation}

The extension of $a_{\rm UV}\to a(r)$ into the bulk is by no means unique.
As a first attempt to do so, we make the substitution
\begin{equation}
\ell\to \ell_{\mathrm{eff}}(r)\equiv\fft1{A'(r)},
\end{equation}
so that
\begin{equation}
a_0(r)=-\fft{\pi^{d/2}}{2\kappa^2(d/2)!^2}\sum_m\alpha_m
(-1)^m\fft{(d+1)!}{(d+1-2m)!(A')^{d+1-2m}}.
\label{eq:a0r}
\end{equation}
This satisfies the requirement that $a(r)$ reproduces the central
charge at fixed points of the flow.  In addition, it incorporates the
metric function at the first derivative level, so that
\begin{equation}
a_0'(r)=-\fft{\pi^{d/2}}{2\kappa^2(d/2)!^2}
\fft{A''}{(A')^d}\sum_m\alpha_m(-1)^{m+1}\fft{(d+1)!(A')^{2m-2}}{(d-2m)!}
\label{eq:a0prime}
\end{equation}
is linear in $A''$.  Following \cite{Freedman:1999gp,Myers:2010xs}, we aim
to demonstrate that $a_0'(r)$ is monotonic by appealing to the bulk equations
of motion.  To do so, we first compute the generalized Einstein tensor
components (\ref{eq:get}) using the curvature components (\ref{eq:riemanns})
for the bulk metric.  The resulting two independent components are
\begin{eqnarray}
G^{t\,(m)}_t&=&-\fft{d!}{2(d-2m)!}(-A'^2)^m+\fft{m(d-1)!}{(d-2m)!}
A''(-A'^2)^{m-1},\nonumber\\
G^{r\,(m)}_r&=&-\fft{d!}{2(d-2m)!}(-A'^2)^m,
\end{eqnarray}
so that
\begin{equation}
G^t_t-G^r_r=A''\sum_mm\alpha_m(-1)^{m+1}\fft{(d-1)!(A')^{2m-2}}{(d-2m)!}.
\end{equation}
Comparison with (\ref{eq:a0prime}) shows that, while the form of $a_0'(r)$
is suggestive, it nevertheless does not match with the difference
$G^t_t-G^r_r$.  However, as indicated above, $a_0(r)$ is not necessarily
unique, and using the Einstein equation as a guide, we now construct a
modified $a$-function which is monotonic.

To proceed, we first note that, at AdS fixed points where the bulk matter
sector contributes vanishing vacuum energy, the background satisfies the
vacuum Einstein equation
\begin{equation}
0=G^r_r\Big|_{\mathrm{AdS}}=\sum_m\left.\alpha_m(-1)^{m+1}
\fft{d!(A')^{2m}}{2(d-2m)!}\right|_{\mathrm{AdS}}.
\end{equation}
(Recall that the cosmological constant term is included in the gravitational
sector through $\alpha_0$.)  This allows us to add a vanishing on-shell
contribution to (\ref{eq:lovfads}), so that
\begin{equation}
f(\mathrm{AdS})=2G_r^r+\sum_m\alpha_mL^{(m)}\Big|_{\mathrm{AdS}}
=\sum_m2m\alpha_m(-1)^m\fft{d!}{(d+1-2m)!\ell^{2m}}.
\end{equation}
In this case, we are led to the definition
\begin{equation}
a(r)=-\fft{\pi^{d/2}}{\kappa^2(d/2)!^2}\sum_mm\alpha_m(-1)^m
\fft{d!}{(d+1-2m)!(A')^{d+1-2m}}.
\label{eq:newla}
\end{equation}
Note that the shift removes the cosmological constant term $\alpha_0$
from the definition of $a(r)$, and matches what is done in constructing
a suitable $a$-function in the leading two-derivative gravity.
This $a$-function can now be seen to satisfy
\begin{equation}
a'(r)=-\fft{d\pi^{d/2}}{\kappa^2(d/2)!^2}\fft{G^t_t-G^r_r}{(A')^d}
=-\fft{d\pi^{d/2}}{(d/2)!^2}\fft{T^t_t-T^r_r}{(A')^d}\ge0,
\end{equation}
where the inequality corresponds to the null energy condition.  Therefore
we have found an appropriate extension of the $a$ central charge which is
indeed monotonic along flows from the UV to the IR.  This result extends
the observation of \cite{Myers:2010xs,Myers:2010tj} that a general holographic
$c$-theorem may be obtained in the presence of higher order corrections,
provided the (linearized) equations of motion remain second order.


\section{$f(R)$ gravity}

In proving the holographic $c$-theorem for Lovelock gravity, we have
constructed the $a(r)$ function (\ref{eq:newla}) entirely out of the
first derivative of the metric function, $A'$.  This ensures that $a(r)$
reproduces the $a$ central charge at fixed points of the flow where
$A'\sim1/\ell$.  However, this construction also guarantees that $a'(r)$
is linear in the second derivative $A''$ so that it may be connected to
the equations of motion.  This connection suggests that having second order
equations is an essential aspect of obtaining the $c$-theorem.  On the
other hand, bulk AdS duals are often considered to be effective theories
where higher derivative corrections naturally arise ({\it e.g.}\ in the
string $\alpha'$ expansion).  Thus it is important to address whether
any holographic $c$-theorem could hold in such higher derivative theories
as well.

Here we take one step towards a fully general investigation by considering
the case of $f(R)$ gravity.  Such theories have been considered from a
cosmological point of view, and are closely related to Brans-Dicke theory
(see {\it e.g.}\ \cite{Sotiriou:2008rp,DeFelice:2010aj,Nojiri:2010wj}).  However, here
we are mainly interested in $f(R)$ gravity as a toy model exhibiting higher
derivative equations of motion.  The action is given by
\begin{equation}
S=\fft1{2\kappa^2}\int d^{d+1}x\sqrt{-g}f(R)+S_{\rm matter},
\end{equation}
where $f(R)$ is a fixed but arbitrary function of the scalar curvature $R$.
The resulting equation of motion is
\begin{equation}
G_{ab}\equiv
FR_{ab}-\ft12fg_{ab}+(g_{ab}\Box-\nabla_a\nabla_b)F=\kappa^2T_{ab},
\label{eq:fReins}
\end{equation}
where
\begin{equation}
F(R)=\fft{df(R)}{dR}.
\end{equation}
Since $F(R)$ is second order in derivatives, the equation of motion is
in general fourth order.

In order to construct an appropriate $a$-function for $f(R)$ gravity, we
follow \cite{Freedman:1999gp,Myers:2010xs} and explore the difference in
the Einstein equation components $G_t^t-G_r^r$.  To proceed, we take the same
metric (\ref{eq:slice}) as used above, and compute the Ricci components
\begin{equation}
R_{\mu\nu}=-(A''+dA'^2)g_{\mu\nu},\qquad
R_{rr}=-d(A''+A'^2),\qquad R=-d(2A''+(d+1)A'^2).
\label{eq:riccis}
\end{equation}
In this case, the Einstein equation (\ref{eq:fReins}) splits into
\begin{eqnarray}
G^\mu_\nu=-[F(A''+dA'^2)+\ft12f-(d-1)A'F'-F'']\delta^\mu_\nu
&=&\kappa^2T^\mu_\nu,\nonumber\\
G^r_r=-F(dA''+dA'^2)-\ft12f+dA'F'&=&\kappa^2T^r_r.
\label{eq:reoms}
\end{eqnarray}
Taking the difference of $G^t_t$ and $G^r_r$ gives
\begin{equation}
G^t_t-G^r_r=(d-1)A''F-A'F'+F''=\kappa^2(T^t_t-T^r_r).
\label{eq:GttmGrr}
\end{equation}
Note that $F$ is a function of $R$, which is in turn a function of $A$
according to (\ref{eq:riccis}).  Thus the higher derivatives of $A$ are
encoded in the $-A'F'+F''$ terms in this equation.

Our aim is to construct a suitable $a(r)$ which reproduces the $a$-anomaly
at the UV boundary and which is subject to a flow governed by
(\ref{eq:GttmGrr}).  We start with the anomaly coefficient itself,
given by (\ref{eq:acc})
\begin{equation}
a_{\mathrm{UV}}=-\fft{\pi^{d/2}}{2\kappa^2}
\fft{\ell^{d+1}}{(d/2)!^2}f(\mathrm{AdS}),
\end{equation}
where this time $f(\mathrm{AdS})$ is simply the on-shell value of $f(R)$ at
the asymptotic AdS fixed point
\begin{equation}
f(\mathrm{AdS})=f(-d(d+1)\ell^{-2}).
\end{equation}
As we saw above, the extension of $a_{\mathrm{UV}}$ to the interior is
not unique.  A straightforward choice would be to replace
$f(\mathrm{AdS})$ by $f(R)$, so that
\begin{equation}
a_0(r)=-\fft{\pi^{d/2}}{2\kappa^2(d/2)!^2}\fft{f(R)}{(A')^{d+1}}.
\label{eq:aguess}
\end{equation}
This is similar to our choice of $a_0(r)$ in (\ref{eq:a0r}), although here
we allow $f(R)$ to contain $A''$ through the dependence on the curvature
scalar.  However, differentiation of $a_0(r)$ with respect to $r$ does not
give any obvious correspondence with the difference (\ref{eq:GttmGrr}).
Thus we seek an improvement to $a_0(r)$, just as we did for the Lovelock
case.

Since we do not want to destroy the matching of the $a$-function with the
actual $a$-anomaly at AdS critical points, we can adjust $a_0(r)$ by at
most functions which vanish at such points.  A natural possibility for
such a function is to take the equations of motion at a critical point.  In
this case, the stress tensor $T_{ab}$ vanishes, and furthermore the
functions $f$ and $F$ become constant.  The $rr$ equation of motion
(\ref{eq:reoms}) then simplifies to
\begin{equation}
G^r_r\Big|_{\rm AdS}=\left[-dA'^2F-\ft12f\right]_{\rm AdS}=0,
\label{eq:adssoln}
\end{equation}
where $A'=1/\ell$.  This suggests that we shift $f(R)$ in (\ref{eq:aguess})
by $2G^r_r$, just as we did in the Lovelock case.  The resulting $a$-function
then takes the form
\begin{equation}
a(r)=\fft{d\pi^{d/2}}{\kappa^2(d/2)!^2}\fft{F(R)}{(A')^{d-1}},
\label{eq:agood}
\end{equation}
where $R$ is given in (\ref{eq:riccis}).  Note that $F(R)$ in the numerator
of this expression is essentially the derivative of the Lagrangian with
respect to $R$ (or equivalently with respect to $R^{tr}{}_{tr}$).  This
suggests a natural connection with an entropy function, as pointed out in
\cite{Myers:2010xs,Myers:2010tj}.  In fact, the appearance of $F(R)$ in the
entropy function for $f(R)$ black holes was initially observed in
\cite{Cognola:2005de}.

With the definition (\ref{eq:agood}), we now see that
\begin{eqnarray}
a'(r)&=&\fft{d\pi^{d/2}}{\kappa^2(d/2)!^2}\fft{-(d-1)A''F+A'F'}{(A')^d}
\nonumber\\
&=&\fft{d\pi^{d/2}}{(d/2)!^2}\fft{-(T^t_t-T^r_r)+F''/\kappa^2}
{(A')^d},
\label{eq:apr}
\end{eqnarray}
where the second line follows from the equation of motion (\ref{eq:GttmGrr}).
If it were not for the $F''$ term, we would then use the null energy condition,
$-(T^t_t-T^r_r)\ge0$, to demonstrate that $a'(r)\ge0$.  This suggests that
the higher derivative nature of $f(R)$ gravity directly impacts the fate of
the holographic $c$-theorem.  In particular, a non-trivial $F''$ contribution
is a direct sign that the gravitational background incorporates up to
four derivatives of $A$.

Taking a step back, it is perhaps not surprising that in this case
monoticity of $a'(r)$ requires not just the weak energy condition on the
matter sector, but also a further condition $F''\ge0$ on the gravity sector.
While we do not see a direct connection with unitarity, it is certainly
plausible that this $F''\ge0$ condition would be related to the absence
of ghost modes in the background of the flow.  One way to investigate this
would be to map $f(R)$ gravity onto Brans-Dicke theory.  In this case,
$F$ plays the role of the Brans-Dicke scalar.  However, it is not clear to
us how $F''$ may be related to any obvious pathologies of the theory.

In searching for a holographic $c$-theorem, we may also need to make a
distinction between perturbative versus non-perturbative expansions in
the higher derivative terms.  For example, at the $R^2$ level, we may
take $f=R+d(d-1)/\ell_0^2+\alpha R^2$, so that $F=2\alpha R$.  The AdS
vacuum with radius $\ell$ is given by the solution to (\ref{eq:adssoln}),
and in this case admits two branches
\begin{equation}
\left(\fft{\ell}{\ell_0}\right)^2=\fft12\pm\sqrt{\fft14-\fft\alpha{\ell_0^2}
\fft{d(d+1)(d-3)}{d-1}}.
\label{eq:adsbkg}
\end{equation}
This may be expanded for small $\alpha$
\begin{eqnarray}
\ell_+&=&\ell_0\left(1-\fft\alpha{2\ell_0^2}\fft{d(d+1)(d-3)}{d-1}
+\cdots\right),\nonumber\\
\ell_-&=&\fft\alpha{2\ell_0}\fft{d(d+1)(d-3)}{d-1}+\cdots.
\end{eqnarray}
We thus see that only the positive branch is smoothly connected to the
finite radius background
in the perturbative limit $\alpha\to0$.  In the context of AdS/CFT, it is
natural to view $\alpha$ as an expansion parameter in an effective theory
with higher curvature interactions.  In this case, we ought to restrict to
only the positive branch.  Perhaps the sign of $F''$ can somehow be attributed
to this choice.  In particular, we have readily found numerical solutions
for this example model coupled to a massless scalar where $a'(r)$ changes
sign (from positive to negative)
along the flow to the IR.  However, all such resulting solutions involve
a domain wall interpolating between the positive and negative branches
of (\ref{eq:adsbkg}), and have the form
\begin{equation}
A(r)\to\begin{cases}r/\ell_+,&r\to\infty\\
|r|/\ell_-&r\to-\infty.\end{cases}
\end{equation}
This `kink up' domain wall solution has the characteristic of a negative
tension wall interpolating between two regions that both open up into AdS
boundaries.  Nevertheless, we have checked that the scalar matter is not
responsible for this negative tension.  Hence, it must have its origin in
the higher derivative gravitational sector.  This at least suggests that
violation of the $c$-theorem is closely related to pathologies in the
gravity sector of the theory that would not arise when treated in a
proper perturbative expansion where the equations of motion can be
perturbatively arranged to use no higher than second order derivatives in
the expansion.

\section{Discussion}

In both the Lovelock and the $f(R)$ case, we have defined the $a$-function
based on a shifted form of the Lagrangian
\begin{equation}
a_{\rm UV}=-\fft{\pi^{d/2}}{2\kappa^2}\fft{\ell^{d+1}}{(d/2)!^2}
f(\mathrm{AdS})\qquad\Rightarrow\qquad
a(r)=-\fft{\pi^{d/2}}{2\kappa^2(d/2)!^2}\fft{f+2G_r^r}{(A')^{d+1}},
\label{eq:genaf}
\end{equation}
where $f=e^{-1}\mathcal L$ is the Lagrangian density in the gravity sector
(not including matter).  Since we take the matter energy density to vanish
at AdS fixed points, the addition of $2G^r_r$ does not affect the
identification of $a(r)$ with the $a$ anomaly coefficient.  However, this
improvement allows $a'(r)$ to be related to the difference of the Einstein
equations $G^t_t-G^r_r$ along the flow.

As noted in \cite{Myers:2010xs}, the $a$-function has a second interpretation
in terms of entanglement entropy.  This can be seen to arise in a natural
manner, provided we perform the $2G^r_r$ shift.  In particular, consider a
general higher curvature action of the form
\begin{equation}
S=\fft1{2\kappa^2}\int d^{d+1}x\sqrt{-g}f(R^{ab}{}_{cd})
+S_{\mathrm{matter}}.
\end{equation}
The corresponding Einstein equation may be written as
\begin{equation}
G_{ab}\equiv F_{(a}{}^{cde}R_{b)cde}
-\ft12fg_{ab}+2\nabla^c\nabla^dF_{acbd}=\kappa^2T_{ab},
\end{equation}
where
\begin{equation}
F_{ab}{}^{cd}=\fft{\delta f(R^{ef}{}_{gh})}
{\delta R^{ab}{}_{cd}}.
\end{equation}
This generalizes the $f(R)$ equation of motion given in (\ref{eq:fReins}).
The higher derivative terms are manifestly present in the Einstein tensor.
However, they vanish on the AdS background where $F_{abcd}$ is
covariantly constant (as it is constructed out of the maximally symmetric
Riemann tensor).  Further taking $T_{ab}$ to vanish in the asymptotic
AdS region, we end up with
\begin{equation}
G^r_r\Big|_{\rm AdS}=\left[-2A'^2F^{r\mu}{}_{r\mu}
-\ft12f\right]_{\mathrm{AdS}}=\left[-2dA'^2F^{tr}{}_{tr}
-\ft12f\right]_{\mathrm{AdS}}.
\end{equation}
The general $a$-function (\ref{eq:genaf}) then takes the form
\begin{equation}
a(r)=\fft{2d\pi^{d/2}}{\kappa^2(d/2)!^2}\fft{F^{tr}{}_{tr}}{(A')^{d-1}}.
\label{eq:afunc}
\end{equation}

If we were to consider black hole entropy in the presence of higher
curvature corrections, it would be natural to use the Wald entropy formula
\cite{Wald:1993nt,Iyer:1994ys,Iyer:1995kg}
\begin{equation}
S=-\fft{2\pi}{2\kappa^2}\int_\Sigma d^{d-1}x\sqrt{-g}
\fft{\delta f}{\delta R_{abcd}}
\epsilon_{ab}\epsilon_{cd}
=\fft{4\pi}{\kappa^2}\int_\Sigma d^{d-1}x\sqrt{h}F^{tr}{}_{tr},
\end{equation}
where the integral is over the area of the horizon with unit binormal
$\epsilon_{ab}$ along $t$ and $r$.  This reduces to the familiar
one-quarter of the horizon area (in $G_N=\kappa^2/8\pi$ units) in the
absence of higher curvatures, where
$F^{ab}{}_{cd}=\fft12\delta^{ab}_{cd}$.
Although this expression is intended to be evaluated at the black hole
horizon, it can nevertheless be generalized into an entropy function
\cite{Goldstein:2005rr,Cremades:2006ke}
\begin{equation}
\tilde C(r)=\fft{4\pi}{\kappa^2}F^{tr}{}_{tr}\sqrt{h}
=\fft{4\pi}{\kappa^2}e^{(d-1)A}F^{tr}{}_{tr},
\label{eq:ctfunc}
\end{equation}
where we have used the explicit form of the metric (\ref{eq:slice}).

Flows of $\tilde C(r)$ have been investigated in the context of the
second law of black hole thermodynamics in higher curvature gravity,
including both Lovelock and $f(R)$ gravity
\cite{Jacobson:1995uq,Cremades:2006ke,Anber:2008js}.  In the case of
$f(R)$ gravity, a $c$-theorem can be proven which generalizes the Hawking
area theorem \cite{Hawking:1971vc} by use of the Raychaudhuri equation
\cite{Jacobson:1995uq,Cremades:2006ke}.  In particular, we consider an
affinely parameterized null congruence given by the tangent vector
$k^a\partial_a=d/d\lambda$ and define
\begin{equation}
\tilde\theta=\fft{d\log\tilde C}{d\lambda}=\theta+k^a\partial_a\log F,
\end{equation}
where $\theta$ is the expansion of the null congruence.  The Raychaudhuri
equation then gives
\begin{equation}
\fft{d\tilde\theta}{d\lambda}
=-\fft1{d-1}\theta^2-\sigma_{ab}\sigma^{ab}
+\omega_{ab}\omega^{ab}-k^ak^bR_{ab}+k^ak^b\nabla_a\nabla_b\log F,
\label{eq:genray}
\end{equation}
and further application of the Einstein equation (\ref{eq:fReins}) reduces
this to
\begin{equation}
\fft{d\tilde\theta}{d\lambda}
=-\fft1{d-1}\theta^2-\sigma_{ab}\sigma^{ab}
+\omega_{ab}\omega^{ab}-\left(\fft{d\log F}{d\lambda}\right)^2
-\fft{\kappa^2k^ak^bT_{ab}}F.
\end{equation}
Provided the congruence is twist-free, and assuming the null energy
condition $k^ak^bT_{ab}\ge0$ along with positivity of $F$, the terms on the
right-hand-side are all non-positive, and as a result we may conclude that
$d\tilde\theta/d\lambda\le0$, which is the statement of the second law
in $f(R)$ gravity \cite{Jacobson:1995uq,Cremades:2006ke}.

For $f(R)$ gravity with the metric written in the explicit form
(\ref{eq:slice}), we define
$k^a\partial_a=-e^{-2A}\partial_t+e^{-A}\partial_r$, in which case
\begin{equation}
\tilde\theta=e^{-A}\left[(d-1)A'+\fft{F'}F\right],
\end{equation}
so that
\begin{eqnarray}
\tilde\theta'&=&e^{-A}\left[-(d-1)A'^2-\left(\fft{F'}F\right)^2
+\fft{(d-1)A''F-A'F'+F''}F\right]\nonumber\\
&=&-e^{-A}\left[(d-1)A'^2+\left(\fft{F'}F\right)^2-\fft{\kappa^2(T^t_t
-T^r_r)}F\right]\le0.
\end{eqnarray}
We see here that, even with a higher order equation of motion, the terms
involving higher derivatives arrange themselves in just the proper manner
to match with the Einstein equation combination (\ref{eq:GttmGrr}).  This
is in contrast with the holographic $a$-function, where $a'(r)$ given in
(\ref{eq:apr}) picks up an additional contribution proportional to $F''$.
Nevertheless, it would be interesting to see if a connection can be made
between the black hole entropy function (\ref{eq:ctfunc}) and the holographic
$a$-function (\ref{eq:afunc}).

Ideally, it would be desirable to construct a general proof of a holographic
$c$-theorem for the $a$-function of (\ref{eq:afunc}) using techniques such
as the generalized Raychaudhuri equation (\ref{eq:genray}).  This would allow
the $c$-theorem to be separated from the particular bulk AdS flow
parameterization of (\ref{eq:slice}).  However, unlike the Raychaudhuri
equation itself, which is independent of dynamics, the incorporation of
$F^{tr}{}_{tr}$ into the generalized expansion $\tilde\theta$ necessarily
brings higher
derivative dynamics into the picture.  Thus it appears unlikely that the
proof of a holographic $c$-theorem for higher curvature gravity can be fully
decoupled from the exact form of the higher order interactions.  (In
fact, we do not expect any $c$-theorem to hold unless additional unitarity
constraints are imposed.)
Nevertheless, we anticipate that it would be possible to make a general
statement in the restricted case where the linearized equations of motion
remain second order in the AdS background.

\begin{acknowledgments}

The idea of a holographic $c$-theorem for $f(R)$ gravity came out of
discussions with R.\ Myers at the Mitchell Institute Application of
AdS/CFT workshop.  JTL also wishes to acknowledge earlier conversations
with C.\ Herzog and D.\ Marolf on AdS $c$-theorems and the Raychaudhuri
equation.  We acknowledge the hospitality of the Khuri lab at
the Rockefeller University, where this work was initiated.
The work of WS was supported in part by the National Science Foundation under
grant number PHY-0903134.
This work was supported in part by the US Department of Energy under grant
DE-FG02-95ER40899.

\end{acknowledgments}


\end{document}